\documentclass[twocolumn,showpacs,preprintnumbers,superscriptaddress,amsmath,amssymb,floatfix,aps]{revtex4}

\usepackage[dvips]{graphicx}
\usepackage{dcolumn}
\usepackage{bm}
\begin{document}


\title{Evanescent character of the repulsive thermal Casimir force}

\author{Vitaly B. Svetovoy}
\affiliation{MESA+ Research Institute, University of Twente, PO 217, 7500 AE Enschede, The Netherlands}

\date{\today}

\begin{abstract}
The physical origin of the negative thermal  correction to the
Casimir force between metals is clarified.  For this purpose  the
asymptotic behavior of the thermal Casimir force is analyzed at
large and small distances in the real frequency representation.
Contributions from propagating and evanescent waves are considered
separately. At large distances they cancel each other in substantial
degree so that only the attractive Lifshitz limit survives. At
smaller separations the repulsive evanescent contribution of
s-polarization dominates in the case of two metals or a metal and a
high-permittivity dielectric. Common origin and order of magnitude
of the repulsion in these two cases demonstrate naturalness of the
controversial large thermal correction between metals.
\end{abstract}

\maketitle

Over the last decade there was a growing interest to the Casimir
force \cite {Cas48} that was measured in a series of recent
experiments \cite{Exper}. Particularly the thermal part of the force
was a subject of active and often controversial discussion (see
\cite{Mil04} for a review). Here we are concentrated on the thermal
force, which does not include the zero point fluctuations. The
repulsion discussed in this paper has the meaning of a negative
thermal correction to the force at zero temperature, but the total
force is always attractive \cite{DLP}. At large distances,
$a\gg\hbar c/T$ ($k_B=1$), the thermal force is given by the
Lifshitz limit \cite{Lif55,DLP,LP8old}
\begin{equation}
F_{Lif}=\frac{T}{16\pi
a^3}\int\limits_0^{\infty}\textrm{d}xx^2\left[\frac{(\varepsilon_{1}+1)(\varepsilon_{2}+1)}
{(\varepsilon_{1}-1)(\varepsilon_{2}-1)}\;e^x-1\right]^{-1},
\label{Lif_lim}
\end{equation}
where $a$ is the distance between parallel plates and $T$ is the
temperature of the system. This formula was derived for two
dielectric plates with the static dielectric constants
$\varepsilon_1$ and $\varepsilon_2$. The force between two ideal
metals can be found from Eq.~(\ref{Lif_lim}) as the limit
$\varepsilon_{1,2}\rightarrow\infty$ that gives
$F_{Lif}=T\zeta(3)/8\pi a^3$. This equation became one of the points
of controversy \cite{Bos00,Therm} because direct calculation of the
thermal force between ideal metals \cite{Meh67} gave the result,
which is two times larger. This contradiction found its resolution
\cite{Hoy03,Sve05}. At large distances only low frequency
electromagnetic (EM) fluctuations contribute to the force. In this
limit s-polarized EM field degenerates to pure magnetic one, which
penetrates freely via a nonmagnetic metal \cite{Sve05}. On the
contrary, p-polarized field is pure electric and reflected by the
metal. For the ideal metal both polarizations are reflected and the
force will be 2 times larger. In this sense the ideal metal is
rather the limit case of a superconductor than of a normal metal
\cite{Ant06}.

The difference between ideal and real metals manifests itself also
at small distances, $a\ll\hbar c/T$. The force between ideal metals
is attractive and small \cite{Meh67}. On the contrary, the force
between real metals is relatively large and repulsive \cite{Bos00}.
This difference did not found yet a clear physical explanation.

In this paper it is demonstrated that at distances $a\lesssim\hbar
c/T$ the evanescent contribution of s-polarization to the thermal
Casimir force dominates for two metals or in the case of a metal and
a high-permittivity dielectric. For both material configurations the
force is repulsive but in the latter case it is free of
controversies accompanying interaction of metals. For ideal metals
the evanescent contribution vanishes but propagating one is
naturally small.

Due to technical convenience in most cases the force is calculated
using the Lifshitz formula \cite{Lif55} written via imaginary
frequencies \cite{DLP,LP8old}. However, in this form a significant
physical information is lost. Originally this formula was presented
in the real frequency domain \cite{LP8old} where the force is
naturally composed of two contributions of the fluctuating fields:
propagating waves (PW) and evanescent waves (EW). It is well known
from optics that each component can be manipulated independently. In
the case of the Casimir force the PW and EW components respond
differently on the change of the material or variation of the
distance. Understanding of this behavior is crucial for tailoring
the force.

Recently Intravaia and Lamrecht \cite {Int05} demonstrated that at
zero temperature, $T=0$, propagating modes give an attractive
contribution to the force between metals while evanescent modes are
responsible for the repulsion. For p-polarization both contributions
are large but cancel each other in substantial degree so that the
resulting force is attractive and small. One might hope to shift the
balance to the repulsion using nanostructured materials \cite
{Int05}. It is not clear for the moment how realistic is this
suggestion but in this paper it is shown that the repulsion can
dominate at least for the thermal part of the force.

The thermal force also revealed interesting cancelations between EW
and PW contributions. At large distances significant cancelations
are realized for dielectrics \cite{Ant06} and metals \cite{Sve06}.
The cancelation is reduced in the non-equilibrium situation when the
temperatures of the bodies are different \cite{Ant06}. It has to be
stressed that in the situation out of equilibrium the real frequency
representation is the only one possible due to absence of
analyticity \cite{Ant06}.

If the wave vector in the vacuum gap is $\mathbf{k}=(q_x,q_y,k_0)$
with $q=\sqrt{q_x^2+q_y^2}$ and $k_0=\sqrt{\omega^2/c^2-q^2}$, then
in the real frequency domain the thermal force is
\cite{LP8old,Ant06} $F=F_{PW}+F_{EW}$, where
\begin{gather}
F_{PW}(a,T)=-\frac{\hbar}{8\pi^2a^3}\textrm{Re}\int_0^{\infty}\frac{d\omega}{e^{\hbar\omega/T}-1}\times\notag\\
\int_0^{\omega/\omega_c}dyy^2\sum_{\mu=s,p}\frac{R^{\mu}(\omega,y)e^{iy}}{1-R^{\mu}(\omega,y)e^{iy}},
\label{FPW}
\end{gather}
\begin{gather}
F_{EW}(a,T)=\frac{\hbar}{8\pi^2a^3}\textrm{Im}\int_0^{\infty}\frac{d\omega}{e^{\hbar\omega/T}-1}\times\notag\\
\int_0^{\infty}dyy^2\sum_{\mu=s,p}\frac{R^{\mu}(\omega,iy)e^{-y}}{1-R^{\mu}(\omega,y)e^{-y}}.
\label{FEW}
\end{gather}
Here $R^{\mu}=r_1^{\mu}r_2^{\mu}$, where $r_{m}^{\mu}$ ($m=1,2$) is
the Fresnel reflection coefficient for body 1 or 2 and transverse
electric ($\mu=s$) or magnetic ($\mu=p$) polarization. The
integration variable $y$ is expressed via the physical values as
$y=2ak_0$ for PW and $y=2a|k_0|$ for EW. The distance dependence is
included in the characteristic frequency $\omega_c=c/2a$. In terms
of $\omega$ and $y$ the reflection coefficients are
\begin{equation}
r^s_{m}=\frac{y-s_{m}}{y+s_{m}},\ \ \
r^p_{m}=\frac{\varepsilon_{m}y-s_{m}}{\varepsilon_{m}y+s_{m}},
\label{rsrp}
\end{equation}
where $s_{m}=\sqrt{\omega^2(\varepsilon_{m}-1)/\omega_c^2+y^2}$ and
$\varepsilon_m=\varepsilon_m(\omega)$ is the dielectric function of
the body $m$.

Consider first the case of two similar metals, which can be
described by the Drude dielectric function
$\varepsilon(\omega)=1-\omega_{p}^2/\omega(\omega+i\omega_{\tau})$,
%
%
where $\omega_{p}$ and $\omega_{\tau}$ are the plasma and relaxation
frequencies, respectively. For the qualitative analysis it will be
assumed that $\omega_{p}\rightarrow\infty,\ \omega_{\tau}\rightarrow
0$ (but both finite). Actual values of the parameters will be taken
into account in numerical calculations.

At large distances, $a\gg\hbar c/T\equiv\lambda_T$ (thermal
wavelength), PW contribution can be found \cite{Ant06} using first
the multiple-reflection expansion
$Re^{iy}/(1-Re^{iy})=\sum_{n=1}^{\infty}R^ne^{iny}$ to avoid poles
and after that one can safely put $R=1$ for both polarizations. The
result will be
\begin{equation}\label{metal_PW}
    F_{PW}(a,T)=\frac{T}{8\pi a^3}\zeta(3)(1+1),
\end{equation}
where we separated two equal contributions from s- and
p-polarizations. Of course, this result coincides with the force
between ideal metals and one could expect that the EW contribution
must be zero. But this is not true.

For s-polarized EW the reflection coefficient is
\begin{equation}\label{rs_metal}
    r^{s}\left( \omega ,iy\right) =\frac{y-\sqrt{\frac{\omega _{p}^{2}}{\omega
_{c}^{2}}\frac{\omega }{\omega +i\omega _{\tau
}}+y^{2}}}{y+\sqrt{\frac{ \omega _{p}^{2}}{\omega
_{c}^{2}}\frac{\omega }{\omega +i\omega _{\tau }} +y^{2}}}.
\end{equation}
If we neglect $y$ in Eq.~(\ref{rs_metal}) in comparison with $\omega
_{p}/\omega _{c}$, then $r^{s}$ will be real and the EW contribution
will vanish. Nonzero result for $F_{EW}^{s}$ originates from very
low frequencies
\begin{equation}
\omega \lesssim \omega _{\tau }\left( \frac{\omega _{c}}{\omega
_{p}}\right) ^{2}.  \label{imp_freq}
\end{equation}
At these frequencies one can expand $e^{\hbar\omega/T}-1\approx
\hbar\omega/T$ in Eq.~(\ref{FEW}). The resulting integral does not
depend on the frequency scale. It means that even arbitrarily small
$\omega_{\tau}$ plays role keeping EW contribution finite
\cite{Sve06}:
\begin{equation}\label{metal_EWs}
    F_{EW}^s(a,T)=-\frac{T}{8\pi a^3}\zeta(3).
\end{equation}
As the result the PW and EW contributions precisely cancel each
other for s-polarization as it happens for dielectrics. It must be
so because a nonmagnetic metal is transparent for s-polarization in
the low frequency limit as was explained above.

For p-polarization very low frequencies (\ref{imp_freq}) are not
important because $r^p$ has significant imaginary part up to
$\omega\sim\omega_{\tau}$. The reflection coefficient is not scale
invariant and the result of integration in Eq.~(\ref{FEW}) will
depend on $\omega_{\tau}$. It is straightforward to show that
$F_{EW}^{p}\sim\omega_{\tau}$ for $\hbar\omega_{\tau}\ll T$ or
$F_{EW}^{p}\sim\omega_{\tau}^{1/2}$ in the opposite limit. In our
qualitative approximation we have to take $F_{EW}^{p}=0$. Therefore,
for p-polarization there is no cancelation between PW and EW
contributions and we recover the result (\ref{Lif_lim}).

It is interesting to see how the result will change at smaller
distances. If the distance is small in comparison with the thermal
wavelength but large in comparison with the penetration depth,
$c/\omega_{p}\ll a\ll\lambda_T$, then the EW contribution will not
change at all. This is because for s-polarization the low
frequencies (\ref{imp_freq}) still dominate and the conclusion on
the behavior of $F^p_{EW}$ also remains true. In contrast, the PW
contribution changes significantly. At small distances there is no
room for standing waves between the plates and the only pressure
which is important is the pressure from the back sides of the plates
due to the black body radiation:
\begin{equation}\label{bb}
    F_{PW}(a,T)=\frac{\pi^2}{90}\frac{T^4}{\hbar^3c^3}(1+1).
\end{equation}
This force is very small and we come to the conclusion that at these
distances the dominant component of the total force is s-polarized
EW:
\begin{equation}\label{Fm_small}
    F(a,T)\simeq -\frac{T}{8\pi a^3}\zeta(3).
\end{equation}
This force is {\it repulsive} and pure {\it evanescent}. To the best
of our knowledge evanescent character of the thermal force at small
distances is stressed here for the first time although one specific
example was demonstrated numerically \cite{Tor04} and rejected as
unacceptable.

\begin{figure}[ptb]
\begin{center}
\includegraphics[width=0.40\textwidth]{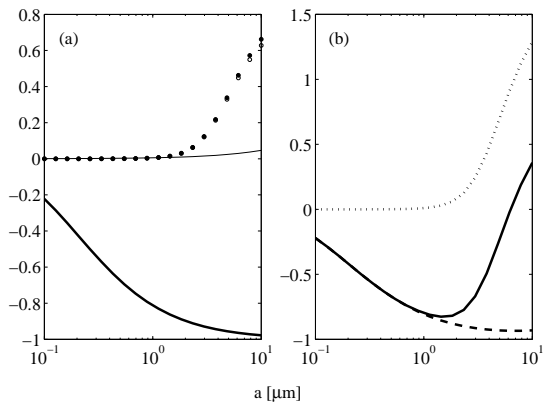}
\vspace{-0.3cm} \caption{(a) Distance dependence for different
components of the thermal force between metals normalized to
$T\zeta(3)/8\pi a^3$. Solid circles correspond to $F_{PW}^s$, open
circles to $F_{PW}^p$, thick line to $F_{EW}^s$, and thin line to
$F_{EW}^p$. (b) Distance behavior of the total force (solid line),
total PW (dotted) and EW (dashed) contributions. } \label{fig1}
\vspace{-0.9cm}
\end{center}
\end{figure}

Of course, Eq.~(\ref{Fm_small}) gives only the leading term and
corrections due to finite $\omega_p$ and $\omega_{\tau}$ are
important for realistic materials but qualitatively the result is
not changed. It is demonstrated in Fig.~\ref{fig1}, which shows
different components of the force in the panel (a) found by
numerical integration of Eqs.~(\ref{FPW}), (\ref{FEW}) at
$T=300^{\circ}\;K$ with the Drude parameters of $Au$:
$\omega_p=9.0\;eV,\; \omega_{\tau}=0.035\;eV$. The total force and
total contributions from PW and EW are shown in panel (b). All
components of the force  are normalized to the natural value
$T\zeta(3)/8\pi a^3$. To get the result in respect with the force at
$T=0$, $\pi^2\hbar c/240a^4$, one has to multiply all the curves to
the factor $1.163a/\lambda_T$.

The repulsive thermal force between metals was discovered by
Bostr\"{o}m and Sernelius \cite{Bos00}. This conclusion was disputed
\cite{Gey03} on the basis that the Lifshitz formula could not be
applicable to metals and the Fresnel reflection coefficients
(\ref{rsrp}) have to be modified at low frequencies. Interaction
between a dielectric and a metal is free of controversies because
$r_1^s r_2^s$ is safely going to zero in the low frequency limit
even if the metal is ideal. But in this case a weak repulsive force
also was demonstrated \cite{Gey05}. Below it is shown that the
repulsion in the system metal-dielectric also originates from the
evanescent s-polarized waves. Moreover, for high-permittivity
dielectrics the repulsion can be of the same order of magnitude as
for metals.

Dielectrics with high permittivity are discussed as the materials
for capacitors and resonators \cite{Tsu06} in GHz region. The
dielectric function for some of them shows no absorption up to a few
THz. For qualitative analysis it will be assumed that in the range
of important frequencies, $\omega\lesssim T/\hbar$, the dielectric
can be described by a constant high permittivity $\varepsilon_2\gg
1$. Frequency dependent dielectric functions of both metal and
dielectric will be used for numerical calculations.

Very low frequencies (\ref{imp_freq}) do not play significant role
in the case of metal and dielectric and for the qualitative analysis
we can take the reflection coefficients of the metal as $r_1^s=-1$
and $r_1^p=+1$. Then the asymptotic behavior of the force at large
and small distances can be found analytically.

At large distances the force is not sensitive to the materials at
all. This is because grazing waves are important in this limit
\cite{Ant06}. For these waves any material is close to a perfect
reflector. For s-polarization $R^s\rightarrow+1$ as for two metals
and we reproduce the result for that configuration:
\begin{equation}\label{s_large}
    F_{PW}^s=\frac{T}{8\pi a^3}\;\zeta(3),\ \ \
    F_{EW}^s=-\frac{T}{8\pi a^3}\;\zeta(3).
\end{equation}
But in contrast with the metals this result is true at much shorter
distances $a\gg\lambda_T\varepsilon_2^{-1/2}$, where
$r^s_2\approx-1$ is a good approximation. The total contribution of
s-polarization is zero as expected. For p-polarization $r^p_2\approx
1$ at the distance $a\gg\lambda_T\varepsilon_2^{1/2}$ that is much
larger than $\lambda_T$. Then
\begin{equation}\label{p_large}
    F_{PW}^p=-\frac{3}{4}\cdot\frac{T}{8\pi a^3}\;\zeta(3),\ \ \
    F_{EW}^p=\frac{7}{4}\cdot\frac{T}{8\pi a^3}\;\zeta(3).
\end{equation}
The factor $-3/4$ in $F_{PW}^p$ originates from the fact that
$R^p\rightarrow -1$ in contrast with the case of s-polarization.
This is a specific property of the metal-dielectric configuration.
For both metal-metal and dielectric-dielectric configurations
$R^p\rightarrow +1$. The total contribution from p-polarization
survives and coincides with the Lifshitz limit (\ref{Lif_lim}) for
$\varepsilon_1\rightarrow\infty$ and $\varepsilon_2\gg1$. It is
interesting to note that the repulsive character of EW is not a
universal property as Eq.~(\ref{p_large}) demonstrates.

The small distance limit is realized for s-polarization at
$a\ll\lambda_T\varepsilon_2^{-1/2}$. In this limit it was found
\begin{equation}\label{s_small}
    F_{PW}^s=-\frac{\pi^2T^4}{90\hbar^3c^3}\frac{3\sqrt{\varepsilon_2}}{2},\ \ \
    F_{EW}^s=-\frac{\pi^2T^4}{90\hbar^3c^3}\varepsilon_2^{3/2}.
\end{equation}
Both contributions are repulsive and large in comparison with the
black body pressure, but EW contribution dominates at
$\varepsilon_2\gg 1$. For p-polarization one can separate the range
of very small distances $a\ll\lambda_T\varepsilon_2^{-3/2}$, but it
has no particular interest because at large $\varepsilon_2$ it
becomes shorter than the penetration depth for the metal,
$c/\omega_p$. Instead, we present the result in the intermediate
range $\lambda_T\varepsilon_2^{-3/2}\ll
a\ll\lambda_T\varepsilon_2^{-1/2}$
\begin{equation}\label{p_small}
    F_{PW}^p=-\frac{\pi^2T^4}{90\hbar^3c^3}\frac{3\sqrt{\varepsilon_2}}{4},\
    F_{EW}^p=\frac{T^2\ln(\varepsilon_2^{3/2}a/\lambda_T)}
    {24a^2\hbar c\sqrt{\varepsilon_2}}.
\end{equation}
%
\begin{figure}[ptb]
\begin{center}
\includegraphics[width=0.40\textwidth]{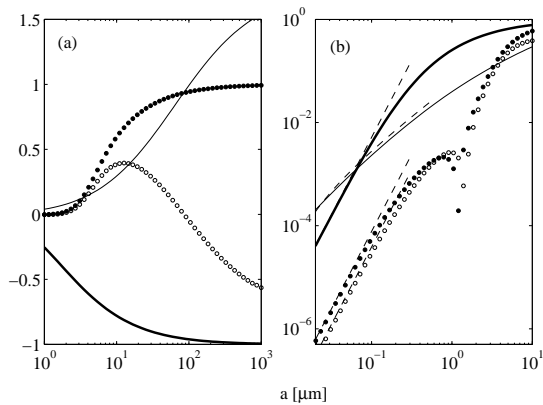}
\vspace{-0.3cm} \caption{Different components of the thermal force
between metal and dielectric. Notations are similar to that in
Fig.~{\ref{fig1}}(a). Panel (a) shows large distance behavior. Panel
(b) shows small distance behavior for absolute values of the forces.
The dashed lines correspond to the asymptotics in
Eqs.~(\ref{s_small}), (\ref{p_small}).} \label{fig2} \vspace{-0.9cm}
\end{center}
\end{figure}

Comparing EW components in (\ref{s_small}) and (\ref{p_small}) in
the distance range $\lambda_T\varepsilon_2^{-3/2}\ll
a\ll\lambda_T\varepsilon_2^{-1/2}$ one can see that
$F_{EW}^p\ll|F_{EW}^s|$. We can conclude that at small distances the
EW contribution from s-polarized field is the largest in the
absolute value and repulsive. When the distance becomes larger
$|F_{EW}^s|$ increases approaching the limit (\ref{s_large}).
$F_{EW}^p$ is also increases but stays much smaller that
$|F_{EW}^s|$ because the former one reaches the asymptotic value
(\ref{p_large}) only at very large distances
$a\gg\lambda_T\varepsilon_2^{1/2}$. When the distance becomes larger
than $\lambda_T\varepsilon_2^{-1/2}$ the PW component in
(\ref{s_large}) becomes important and $F_{EW}^s$ is not dominant any
more. It means that the total force has to have a minimum where it
is repulsive.

Figure \ref{fig2} shows the numerical results for different
components of the thermal force. The calculations were done for the
ideal metal and dielectric with the frequency independent
permittivity $\varepsilon_2=100$ at $T=300^{\circ}\;K$. Panel (a)
shows behavior at large distances. An important result of the
numerical analysis is that the large distance limit is realized at
$a$ approximately 100 times larger than
$\lambda_T\varepsilon_2^{-1/2}$ for s-polarization or
$\lambda_T\varepsilon_2^{1/2}$ for p-polarization. As one can see
for p-polarization the PW and EW components did not fully reach the
asymptotic values -3/4 and 7/4 even at $a=1000\;\mu m$. Nonmonotonic
behavior of $F^p_{PW}$ also should be mentioned. At smaller
distances the absolute values of the PW and EW components are shown
in panel (b). Dashed lines represent the asymptotics (\ref{s_small})
and (\ref{p_small}).

\begin{figure}[ptb]
\begin{center}
\includegraphics[width=0.40\textwidth]{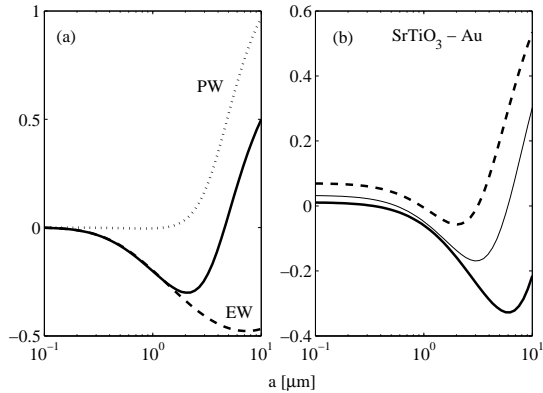}
\vspace{-0.3cm} \caption{(a) Total PW (dotted), total EW (dashed)
and the total force (solid) for the ideal metal and ideal dielectric
with $\epsilon_2=100$. (b) The total force between realistic
materials: $Au$ and $SrTiO_3$, $\epsilon_2(0)=320$. Dashed, thin
solid, and thick solid lines correspond to $T=300, 200, {\rm and}
100^{\circ}\;K$, respectively. } \label{fig3} \vspace{-0.9cm}
\end{center}
\end{figure}

Figure \ref{fig3}(a) shows the total PW and EW contributions
together with the total force. It demonstrates the expected minimum
of the total force. Panel (b) shows the total force between
$SrTiO_3$ single crystal \cite{Tsu06} and $Au$. Frequency dependence
of the dielectric functions of both materials was taken into
account. The results are presented for three different temperatures.
Influence of the temperature is significant because the absorption
resonances of $SrTiO_3$ are somewhat below the room temperature.
Frequency dependence of $\varepsilon_2$ increases both EW
components, thus reducing the repulsive force. Of course, with the
temperature decrease the minimum becomes deeper and its position is
shifted to larger distances.

In conclusion, we analyzed the PW and EW contributions to the
thermal Casimir force between two metals and a metal and a
high-permittivity dielectric. For both material configurations the
repulsive s-polarized EW contribution dominates at $a\lesssim
\lambda_T$. The thermal repulsion between metals was disputed in the
literature but the result of this paper demonstrates that similar
effect is realized for uncontroversial case of metal and dielectric.


\end{document}